\documentclass{sig-alternate}
\DeclareMathOperator*{\argmax}{arg\,max}

\usepackage{ctable}
\usepackage{multirow}
\usepackage{balance}
\usepackage{booktabs}
\usepackage{hyphsubst}
\usepackage{graphicx}
\usepackage{amsmath}
\usepackage{url}
\usepackage{hhline}
\usepackage[caption = false]{subfig}

\permission{Copyright is held by the International World Wide Web Conference Committee (IW3C2). IW3C2 reserves the right to provide a hyperlink to the author's site if the Material is used in electronic media.}
\conferenceinfo{WWW'15 Companion,}{May 18--22, 2015, Florence, Italy.} 
\copyrightetc{ACM \the\acmcopyr}
\crdata{xxx-x-xxxx-xxxx-x/xx/xx. \\
http://dx.doi.org/xx.xxxx/xxxxxxx.xxxxxxx}

\begin{document}

\title{Attention Please! A Hybrid Resource Recommender Mimicking Attention-Interpretation Dynamics}

\numberofauthors{6}
\author{
\alignauthor
Paul Seitlinger\\
       \affaddr{Graz University of Technology\\Graz, Austria}\\
       \email{paul.seitlinger@tugraz.at}
\alignauthor
Dominik Kowald\\
       \affaddr{Know-Center\\Graz, Austria}\\
       \email{dkowald@know-center.at}
\alignauthor
Simone Kopeinik\\
       \affaddr{Graz University of Technology\\Graz, Austria}\\
       \email{simone.kopeinik@tugraz.at}
\and
\alignauthor
Ilire Hasani-Mavriqi\\
       \affaddr{Graz University of Technology\\Graz, Austria}\\
       \email{ihasani@know-center.at}
\alignauthor
Tobias Ley\\
       \affaddr{Tallinn University\\Tallinn, Estonia}\\
       \email{tley@tlu.ee}
\alignauthor
Elisabeth Lex\\
       \affaddr{Graz University of Technology\\Graz, Austria}\\
       \email{elisabeth.lex@tugraz.at}
}

\maketitle
\begin{abstract}
Classic resource recommenders like Collaborative Filtering (CF) treat users as being just another entity, neglecting non-linear user-resource dynamics shaping attention and interpretation. In this paper, we propose a novel hybrid recommendation strategy that refines CF by capturing these dynamics. The evaluation results reveal that our approach substantially improves CF and, depending on the dataset, successfully competes with a computationally much more expensive Matrix Factorization variant.
\end{abstract}



\section{Introduction} \label{sec:intro}
The Web features a huge amount of data and resources that are potentially relevant and interesting for a user. Users are yet often unable to evaluate all available alternatives due to cognitive limitations of their minds. Recommender systems have been proved as being a valid approach for Web users to cope with the information overload \cite{kantor2011recommender} -- with Collaborative Filtering (CF) being one of the most successful methods \cite{bar2013improving}. CF aims to recommend resources to a user based on the digital traces she leaves behind on the Web, i.e., her interactions with resources and the interactions of other similar users.

Recent advances in the interdisciplinary field of Web Science provide even more comprehensive digital traces of social actions and interactions that can be exploited in recommender research. At least implicitly, recommender research has implemented interesting assumptions about structures and dynamics in Social Information Systems (SIS), such as MovieLens, LastFM or BibSonomy. For instance, by computing matrices or high-dimensional arrays, approaches like CF represent and process SIS as networks or graphs, which relate entities of different quality (e.g., users, resources, time, ratings, tags, etc.) to each other. That way, a compositional view is taken that reminds of a material-semiotic perspective (e.g., \cite{law2009actor}), assuming that we gain a deeper understanding of the intention or function of an entity, if we consider the associations it has established with other entities. Put differently, ``everything in the social and natural worlds [is regarded] as a continuously generated effect of the webs of relations within which they are located'' (\cite{law2009actor}, p. 142).

However, if we look at the machinery underlying CF, it becomes clear that structurally, the algorithm treats users as being just another entity, such as a tag or a resource. We regard this indifference as a structuralist simplification abstracting from individuals' complexity. The structuralist stance also runs the risk of neglecting nonlinear, dynamic processes going on between different entities, such as a user's intentional state (e.g., attentional focus, interpretations, decision making) and resources (articles, movies, etc.) consumed in the past.

It is the main goal of this work to take a closer look at these dynamics and to capture them by means of an appropriate model. Each user develops subjectivity, an idiosyncratic way of perceiving and interpreting things in the world, which manifests itself in particular preferences. Partially, this development evolves through a user's trajectory in the SIS (e.g., \cite{fu2012collaborative}). Every resource that we decide to collect corresponds to a learning episode \cite{dennerlein2014making}: Depending on the resource's attributes, the episode causes a shift in attention, particularly, in attentional tunings for certain attributes as well as a shift in mental categories (conceptual clusters), which influences our decision-making (e.g., \cite{love2004sustain}). The shape that mental patterns (e.g., attentional tunings and conceptual clusters) acquire, is governed by both, the environment and the current mental state. The acquired pattern in turn orients the user toward particular resources and hence, closes the loop of the environment-user dynamics.

\subsection{A Model of Attention-Interpretation Dynamics}
The connectionist approach towards human categorization (e.g., \cite{rumelhart1988learning,kruschke1992alcove,love2004sustain}) seems to be the method of choice to model these dynamics: Input, hidden and output units are interconnected within a multi-layer network mapping inputs (e.g., resource attributes) to outputs (e.g., decision to take or leave a resource). Tunings of particular units (e.g., the attentional weight of an input unit) and of the interconnections evolve as the network encounters new exemplars (e.g., resources). A particularly flexible and self-supervising model is SUSTAIN \cite{love2004sustain}, whose number of hidden units (i.e., clusters) is not chosen in advance but incrementally discovered in the course of the learning trajectory. It adapts to the trajectory's complexity and only recruits a new cluster if the current resource cannot be assimilated with the already existing clusters.

That way, SUSTAIN can be applied to do more justice to a user's way of interacting within an SIS, e.g., in terms of the resources the user attends to and of the semantic interpretation of the resources. For instance, consider a user exhibiting a broad attentional focus because the collected resources cover a broad range of topics. Modeling that user's trajectory by means of SUSTAIN will yield a network with a uniform distribution of attentional tunings (i.e., for different topics) and a number of distinct clusters since the resources fall inside different regions of the topic space. On the other hand, consider a user with a narrow attentional focus that corresponds to a list of resources addressing similar topics. The result will be a peaky attentional distribution and a small number of clusters.

Since SUSTAIN is able to differentiate between users with respect to attention and interpretation, we assume that it can be applied to anticipate user-specific preferences and decisions on resources. In this work, we therefore introduce a resource recommender that draws on SUSTAIN to model a user's traces (e.g., posts) in form of a connectionist network representing the user's attentional focus and semantic clusters. Given the traces provide sufficient information for training the network, we assume that a recommender equipped with SUSTAIN can be applied to simulate a user's decision making on a set of resources and therefore, improve the recommender's accuracy. For a first proof of concept, we proceed in two steps: First, we make use of CF to gather a more restricted set of resources. Second, SUSTAIN simulates how the user interacts with each of the resources and simulates the resulting decision of the user to take or leave the resource. These two steps constitute our novel hybrid resource recommender approach termed SUSTAIN+CF$_U$ (see Section \ref{sec:approach}). In brief, our research question of this work is:

\textit{\textbf{RQ:} Do resource recommendations become more accurate if a set of resources identified by CF is processed by SUSTAIN to simulate user-specific attentional and conceptual processes?}

The remainder of this paper is organized as follows: In Section \ref{sec:relwork}, we consider related work that has inspired our own approach, whose algorithm is described in Section \ref{sec:approach}. In Section \ref{sec:meth}, we describe our methodology applied to compare the performance of our SUSTAIN+CF$_U$ approach with several baseline algorithms. In Sections \ref{sec:results} and \ref{sec:conclusion}, we then report our evaluation results and provide a conclusion and outlook, respectively.

\section{Related Work} \label{sec:relwork}
At the moment, we identify three main research disciplines that are related to our work.

\textbf {Collaborative Filtering Extensions:} 
One of our previous studies in this field \cite{lacic2014recommending}, introduces the so-called 
\textit{Collaborative Item Ranking Using Tag and Time Information (CIRTT)} approach, which extends CF in social
tagging systems by incorporating tag and time information. This approach combines
user-based and item-based CF with the information of tag frequency and recency by applying the base-level learning (BLL) equation coming
from human memory theory.

An extensive survey on CF was recently conducted by \cite{Shi2014Survey}. In this survey, the authors classify CF approaches based on the type of information that is processed and
the type of paradigm applied. Furthermore, CF extensions are defined as approaches which enrich
classic CF algorithms with valued side information on users and resources. Analogous categorization of
CF studies is performed in \cite{Adomavicius2005Survey} as well. Additionally, these studies have identified challenges that are crucial to
the future research on CF. In this context, authors state the fact that there is a lack of
studies which address issues on recommender systems from the psychological perspective. To the best of our knowledge, there has been no remarkable endeavors which tend to combine
the implementation of a dynamic and connectionist model of human cognition, such as SUSTAIN, with
existing CF algorithms.

\textbf{Recommender Systems and User Modeling:} 
The work of \cite{Cremonesi2012Study} distinguishes between recommender systems that provide
non-personalized and personalized recommendations. Whereas, non-personalized recommender systems are not based on
user models, personalized ones choose resources by considering the user profile (e.g. previous user
interactions or user preferences). Considerable amount of techniques have been proposed to design the user model in terms of resource recommendations \cite{Jawaheer2014Usermodeling} \cite{Coleho2010Usermodeling}. 
Among them, some approaches aim to provide dynamically adapted personalized recommendations to users \cite{Dooms2013Hybrid}.

A specific research topic, which is increasingly gaining popularity, is human decision making in
recommender systems \cite{Chen2013DecisionStudy}. The work presented in \cite{Cremonesi2012Decision}
systematically analyzes recommender systems as decision support systems based on the nature of users' goals and the
dynamic characteristics of the resource space (e.g., availability of resources).
However, there is still lack of research on investigating user decision processes at a detailed
level and considering integrating scientific facts from psychology. 
Thus, we consider that our proposed approach contributes to this line of research.

\textbf{Long Tail Recommendations and User Serendipity: }
Another concept within the area of recommender systems, that needs to be more extensively considered in the future, 
is termed as long tail recommendations. Basically, the long tail refers to the
resources that have low popularity \cite{Shi2014Survey}. It is of a huge interest to show how
recommendations of these long tail resources can impact user satisfaction. Furthermore, it is important to
investigate if additional revenue can be generated by the recommender systems from the long tail resources \cite{Shi2014Survey} \cite{Yin2012Longtail} \cite{ShiLei2013Longtail}.

In this line of research, various solutions have been proposed to overcome the problem of over-specialization and concentration-bias in recommender systems \cite{Adamopoulos2014Specialization}.
The problem of concentration-bias becomes evident since traditional CF algorithms recommend resources based on a the users' previous history of activities. 
Hence, resources with the most occurrences in this history are typically repeatedly recommended to users causing narrowing the choices, by excluding other resources which might be of interest.
Additionally, recommending resources based on user's previous activities or preferences yields to over-specialization of recommendations.
However, balancing between information overload and facilitating users to explore new horizons by recommending serendipitous choices, remains a challenge to be considered for future research.


\section{Approach} \label{sec:approach}
\begin{table}[t!]
  \setlength{\tabcolsep}{2.8pt}	
  \centering
    \begin{tabular}{ll|l}
    \specialrule{.2em}{.1em}{.1em}
      Function    & Symbol & Value\\\hline 
      Attentional focus  & $r$ & 9.998\\\hline
      Cluster competition  & $\beta$ & 6.396\\\hline
      Learning rate  & $\eta$ & .096\\\hline
      Threshold	  & $\eta$ & .5 \\\hline
		\specialrule{.2em}{.1em}{.1em}								
    \end{tabular}
    \caption{SUSTAIN's best fitting parameters for unsupervised learning as suggested in \cite{love2004sustain}.}	
  \label{tab:parameters}
\end{table}

In this section, we introduce a novel hybrid resource recommender approach termed SUSTAIN+CF$_U$, which further personalizes and improves user-based Collaborative Filtering (CF$_U$). SUSTAIN \cite{love2004sustain}, briefly described in the introductory section, is a computational model of human category learning. It builds upon the assumption that things can be described by patterns of correlated features, that form $n$-dimensional clusters, and attentional tunings representing each dimension's relative importance. To apply this theory when recommending Web resources, we use 500 LDA topics derived from the tag assignments as the $n$ features \cite{griffiths2007topics} (see Section \ref{sec:datasets}). 
On the basis of the resources a user has bookmarked in the past (i.e., the training set of a user), each user's personal attentional tunings and cluster representations are created in a training phase. Following an unsupervised 
learning procedure, we start simple, with one cluster and expand the number of clusters if necessary. Please note, that all SUSTAIN-specific parameter settings are adopted from \cite{love2004sustain} (see Table \ref{tab:parameters}). 

For each resource in the training set of user $u$, we start by calculating the distance $\mu_{ij}$ to cluster $j$ at dimension $i$ as described in equation \eqref{eq:four}:

\begin{align}\label{eq:four}
    \mu_{ij} = I^{pos_i} - H_j^{pos_i}
\end{align} 
where $I$ is the $n$-dimensional input vector, which represents the topics of this resource, and vector $H_j$ is cluster $j$'s position in the $n$-dimensional feature space, which holds a value for each topic and is initially set to $\vec{0}$. In the next step, the distance $\mu_{ij}$ is used to calculate the activation value $H_j^{act}$ of the $j^{th}$ cluster by equation \eqref{eq:five}:

\begin{align}\label{eq:five}
   H_j^{act} = \frac{\sum_{i=1}^{n}{(\lambda_i)^re^{-\lambda_i\mu_{ij}}}}{\sum_{i=1}^{n}({\lambda_i})^r}
\end{align} 
where $\lambda_i$ represents the tuning (weight) of dimension $i$ and acts as a multiplier on $i$ in calculating the activation. Initially, vector $\lambda$ is set to $\vec{1}$ 
and evolves during the training phase according to equation \eqref{eq:thirteen} calculated at the end of every training iteration (i.e., after including a resource). 
$r$, which is set to 9.998, is an attentional parameter that accentuates the effect of $\lambda_i$: if $r$ = 0, all dimensions are weighted equally.  
Clusters compete with one another, thus weakening the most activated one (i.e., $max(H_j^{act})$). This form of lateral inhibition is calculated by equation \eqref{eq:twelve}, 
which gives an activation value $H_m^{out}$ 
for the winning cluster $m$: 
\begin{align}\label{eq:twelve}
    H_m^{out} = \frac{(H_m^{act})^\beta}{\sum_{i=1}^{n}{(H_i^{act})^\beta}}H_m^{act}
\end{align} 
where $\beta$ is the lateral inhibition parameter and is set to 6.396. If the activation value $H_{m}^{out}$ of the most activated (i.e., winning) cluster is beneath a given threshold $\tau$ = .5, a new cluster is created, representing the topics of the currently processed resource. At the end of an iteration, the tunings of vector $\lambda$ are updated given by equation \eqref{eq:thirteen}:
\begin{align}\label{eq:thirteen}
   \Delta \lambda_i = \eta^{-\lambda_i\mu_{im}}(1 - \lambda_i\mu_{im})
\end{align} 
where $j$ indexes the winning cluster and the learning rate $\eta$ is set to .096. In a final step, the position vector of the winning cluster, which holds a value for each of the $n$ topics, is recalculated as described by equation \eqref{eq:finaleq}:
\begin{align} \label{eq:finaleq}
   \Delta H_m^{pos_i} = \eta(I^{pos_i} - H_m^{pos_i})
\end{align} 
The training phase is completed when steps \eqref{eq:four} to \eqref{eq:finaleq} were subsequently processed for every resource in a user's training set. For each user, this results in a particular vector of attentional tunings 
$\lambda$ and a set of $j$ cluster vectors $H_j$.

We make use of this user-specific pattern to calculate personalized resource recommendations by incorporating CF$_U$ into our hybrid approach SUSTAIN+CF$_U$. 
Thus, first, we determine the top-$100$ resources identified by CF$_U$ (see Section \ref{sec:algos}) as a candidate set C$_u$ of potentially relevant resources for the target user $u$. Then, for each candidate $c$ in C$_u$, we calculate $H_m^{out}$ following equations \eqref{eq:four} to \eqref{eq:twelve}. $H_{m}^{out}(c)$ can then be added to CF$_U(c)$ as shown in equation \eqref{eq:hybrid} in order to determine the set of $k$ recommended resources $RecRes(u)$ for user $u$:
\begin{align} \label{eq:hybrid}
   RecRes(u) = \argmax_{c \in C_u}^{k}(\underbrace{{\alpha}H_{m}^{out}(c) + {(1 - \alpha)}CF_U(c)}_{SUSTAIN+CF_U})
\end{align} 
where $\alpha$ can be used to inversely weigh the two components of our hybrid approach. For now, we set $\alpha$ to .5 in order to equally weight SUSTAIN and CF$_U$.

Our approach as well as the baseline algorithms described in Section \ref{sec:algos} (but for WRFM) and the evaluation method described in Section \ref{sec:eval} are completely implemented in Java within our \textit{TagRec} recommender benchmarking framework \cite{domi2014a}, which is freely available via GitHub\footnote{\url{https://github.com/learning-layers/TagRec/}}.

\section{Methodology} \label{sec:meth}
In this section we describe the methodology used in this work, including the datasets, the evaluation method and metrics and the baseline algorithms.

\subsection{Datasets} \label{sec:datasets}
For the evaluation of our approach, we used datasets gathered from well-known social tagging systems. We focused on social tagging systems for our study not only because their datasets are freely-available for scientific purposes but also because tagging data can be easily utilized to derive semantic topics for resources from it (that are needed for our approach) \cite{griffiths2007topics} by means of Latent Dirichlet Allocation (LDA) \cite{krestel2009tag}. LDA is a probability model that helps to find latent semantic topics for documents (i.e., resources). In the case of tagging data, the model takes the assigned tags of the resources as input and returns the identified topic distributions for each resource. We implemented LDA using the Java framework Mallet\footnote{\url{http://mallet.cs.umass.edu/}} with Gibbs sampling and 2000 iterations as suggested in the framework documentation and by related work (e.g., \cite{krestel2009latent}). Moreover, we set the number of latent topics to 500 (see also \cite{kintsch2011construction,paul2013}) and considered only identified topics for resources that have a minimum probability value of .01 in order to reduce noise and to meaningfully limit the number of assigned topics.

The statistics of our chosen datasets are shown in Table \ref{tab:datasets}. We focused on BibSonomy\footnote{\url{http://www.kde.cs.uni-kassel.de/bibsonomy/dumps/}}, CiteULike\footnote{\url{http://www.citeulike.org/faq/data.adp}} and Delicious\footnote{\url{http://files.grouplens.org/datasets/hetrec2011/hetrec2011-delicious-2k.zip}} to test our approach in three different settings that differ in their dataset sizes. To reduce computational effort, we randomly selected 20\% of the CiteULike user profiles \cite{gemmell2009improving} (the other datasets were processed in full size). We did not use a $p$-core pruning approach to avoid a biased evaluation (see \cite{Doerfel2013}) but excluded all unique resources, i.e., resources that have only been bookmarked once (see \cite{parra2010improving}).
 
\begin{table}[t!]
  \setlength{\tabcolsep}{2.6pt}	
  \centering
    \begin{tabular}{l|lllll}
    \specialrule{.2em}{.1em}{.1em}
      Dataset			& $|P|$			& $|U|$	& $|R|$	& $|T|$	& $|TAS|$	 							\\\hline 
      BibSonomy	  & 82,539 & 2,437  & 28,000		& 30,919	& 339,337				\\\hline
			CiteULike		& 105,333 & 7,182  & 42,320		& 46,060	& 373,271				\\\hline
			Delicious		& 59,651 & 1,819  & 24,075		& 23,984	& 251,542			\\\hline
		\specialrule{.2em}{.1em}{.1em}								
    \end{tabular}
    \caption{Properties of the datasets, where $|P|$ is the number of posts, $|U|$ is the number of users, $|R|$ is the number of resources, $|T|$ is the number of tags and $|TAS|$ is the number of tag assignments.}	
  \label{tab:datasets}
\end{table}

\subsection{Evaluation Method and Metrics} \label{sec:eval}
In order to evaluate our algorithm and to follow common practice in recommender systems research (e.g., \cite{huang2014utilizing,zheng2011recommender}), we split our datasets into training and test sets. Therefore, we followed the method described in \cite{lacic2014recommending} to retain the chronological order of the posts. This also simulates well a real-world environment, where future interactions are tried to be predicted based on interactions in the past \cite{campos2013time}. Hence, for each user we used her 20\% most recent posts for testing and the rest for training.

To finally determine the performance of our approach as well as of the baseline methods, we compared the top-$20$ recommended resources determined by each algorithm for a user with the relevant resources 
in the test set using a variety of well-known evaluation metrics \cite{ParraSahebi,Herlocker2004} in recommender systems research. In particular, we report Normalized Discounted Cumulative Gain (nDCG@20),
Mean Average Precision (MAP@20), Recall (R@20) and Precision (P@20). Moreover we show the performance of the algorithms for different numbers of recommended resources ($k = 1 - 20$) by means of Precision / Recall plots.

\begin{table*}[t!]
\begin{center}
\begin{tabular}{l|l|llll|ll}
\specialrule{.2em}{.1em}{.1em}
  Dataset & Metric &  MP 	 	& CF$_{R}$ 	& CB$_{T}$ & WRMF 										& CF$_{U}$					&SUSTAIN+CF$_{U}$	\\\hline
	
  \multirow{4}{*}{\centering{\centering{BibSonomy}}}  
  &  nDCG@20 		&	$.0142$ 	 		& $.0569$ & $.0401$ & $.0491$ 										&	$.0594$						& $.\textbf{0665}$\\ 
  &  MAP@20 		&	$.0057$ 	 		& $.0425$ & $.0211$ & $.0357$ 										&	$.0429$						& $.\textbf{0492}$\\  
  &  R@20	 			&	$.0204$ 	 		& $.0803$ & $.0679$ & $.0751$ 										&	$.0780$						& $.\textbf{0859}$\\  
  &  P@20 			&	$.0099$ 	 		& $.0223$ & $.0272$ & $.0132$ 										&	$.0269$						& $.\textbf{0290}$\\\hline
	
  \multirow{4}{*}{\centering{\centering{CiteULike}}}   
  &  nDCG@20 		&	$.0064$ 	 	& $.\textbf{1006}$ & $.0376$ & $.0411$  						&	$.0753$						& $.0857$\\ 
  &  MAP@20 		&	$.0031$ 	 	& $.\textbf{0699}$ & $.0170$ & $.0210$  						&	$.0468$						& $.0555$\\   
  &  R@20 			&	$.0090$ 	 	& $.\textbf{1332}$ & $.0697$ & $.0658$  						&	$.1149$						& $.1251$\\    
  &  P@20 			&	$.0023$ 	 	& $.\textbf{0289}$ & $.0174$ & $.0218$  						&	$.0257$						& $.0280$\\\hline

  \multirow{4}{*}{\centering{\centering{Delicious}}}   
  &  nDCG@20 		&	$.0038$ 	 	& $.1148$ & $.0335$ & $.\textbf{1951}$  						&	$.1327$						& $.1633$\\ 
  &  MAP@20 		&	$.0011$ 	 	& $.0907$ & $.0134$ & $.\textbf{1576}$  						&	$.0949$						& $.1156$\\   
  &  R@20 			&	$.0071$ 	 	& $.1333$ & $.0447$ & $.\textbf{2216}$  						&	$.1662$						& $.1946$\\    
  &  P@20 			&	$.0017$ 	 	& $.0512$ & $.0173$ & $.\textbf{1229}$  						&	$.0843$						& $.0982$\\\hline
									
	\specialrule{.2em}{.1em}{.1em}
\end{tabular}
\vspace{-2mm}
\caption{nDCG@20, MAP@20, R@20 and P@20 estimates for BibSonomy, CiteULike and Delicious showing that our proposed approach SUSTAIN+CF$_U$ not only outperforms CF$_U$ in all settings but also is able to successfully compete with WRMF (\textit{Note}: highest accuracy values per dataset over all algorithms are highlighted in bold).}
 \label{tab:full_norm}
\end{center}
\vspace{-6mm}
\end{table*}

\subsection{Baseline Algorithms} \label{sec:algos}
We selected a set of well-known resource recommender baseline algorithms in order to determine the performance of our novel approach in relation to these approaches. Hence, we have not only chosen algorithms that are similar to our approach in terms of their processing steps (CF$_U$ and CB$_T$) but also current state-of-the-art methods for personalized resource recommendations (CF$_R$ and WRMF) along with a simple unpersonalized approach (MP).

\textbf{Most Popular (MP):}
The most simple method we compare our algorithm to, is the \textit{Most Popular (MP)} approach that ranks the resources by their total frequency in all posts \cite{ParraSahebi}. In contrast to the other chosen baselines, the MP approach is non-personalized and thus, recommends for any user the same set of resources.

\textbf{User-based Collaborative Filtering (CF$_U$):}
\textit{User-based Collaborative Filtering (CF$_U$)} typically consists of two steps. First, the most similar users (the so-called $k$ nearest neighbors) for a target user are found based on a specific similarity measure and second, the resources of these neighbors, that are new to the target user, are suggested. This procedure is based on the idea, that if two users had a similar taste in the past, they will probably share the same taste in the future and thus, will like the same resources \cite{schafer2007collaborative}. We calculated the user-similarities based on the binary user-resource matrix and using the cosine-similarity measure (see \cite{zheng2011recommender}). Moreover, we set the neighborhood size $k$ to 20 as suggested in \cite{gemmell2009improving} for CF$_U$ in social tagging systems.

\textbf{Resource-based Collaborative Filtering (CF$_R$):}
In contrast to CF$_U$, \textit{Resource-based Collaborative Filtering (CF$_R$)} (also known as Item-based CF), identifies potentially interesting resources for a user via computing similarities between resources instead of similarities between users. Hence, this approach processes the resources a user has bookmarked in the past in order to find similar resources to be recommended \cite{sarwar2001item}. As in the case of CF$_U$, we calculated similarities based on the binary user-resource matrix using cosine similarity and focused on a resource-neighborhood size $k$ of 20 \cite{zheng2011recommender,gemmell2009improving}.

\textbf{Content-based Filtering using Topics (CB$_T$):}
Content-based filtering (CB) methods recommend resources to users based on a comparison between the resource content and the user profile \cite{basilico2004unifying}. Hence, this approach does not need to calculate similarities between users or resources (as done in CF methods) but directly tries to map resources and users. We implemented this method in the form of \textit{Content-based Filtering using Topics (CB$_T$)} since topics are the only content-based features available in our social tagging datasets (see Section \ref{sec:datasets}). The similarity between the topic vector of a user and a resource has been calculated using the cosine similarity measure.

\textbf{Weighted Regularized Matrix Factorization\\(WRMF):}
\textit{WRMF} is a model-based recommender method for implicit data (e.g., posts) based on the state-of-the-art Matrix Factorization (MF) technique. MF factorizes the binary user-resource matrix into latent user- and resource-factors, that represent these entities, in a common space. This representation is used to map resources and users and thus, find resources to be recommended for a specific user. WRMF defines this task as a regularized least-squares problem based on a weighting matrix, which differentiates between observed and unobserved activities in the data \cite{hu2008collaborative}. The results for WRFM presented in Section \ref{sec:results} have been calculated using the MyMediaLite framework\footnote{\url{http://www.mymedialite.net/}} with 500 latent factors, 100 iterations and a regularization value of .001.

\section{Results} \label{sec:results}
\begin{figure*}[t!]
   \centering 
   \subfloat[BibSonomy]{ 
      \includegraphics[width=0.33\textwidth]{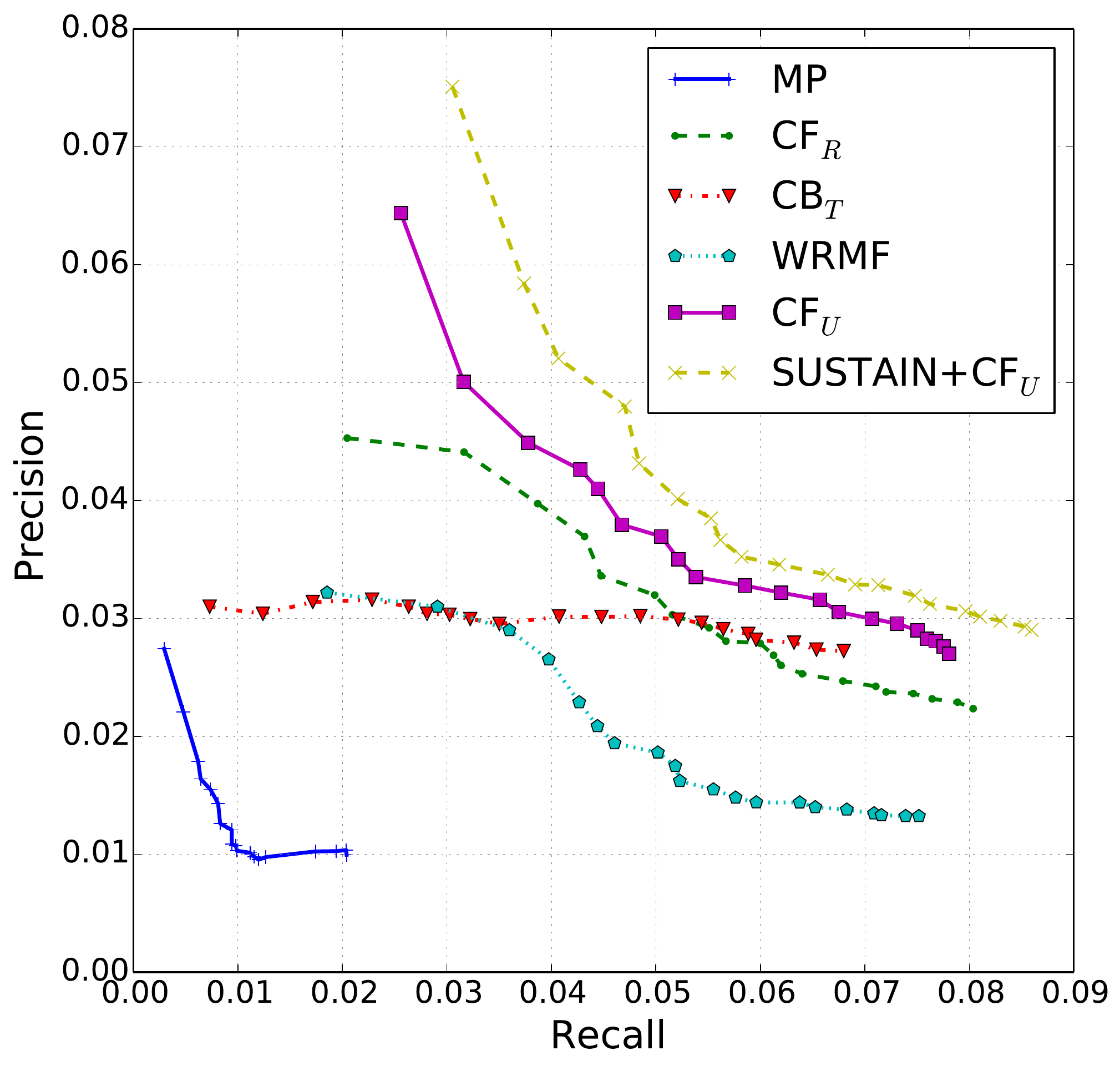} 
   } 
   \subfloat[CiteULike]{ 
      \includegraphics[width=0.33\textwidth]{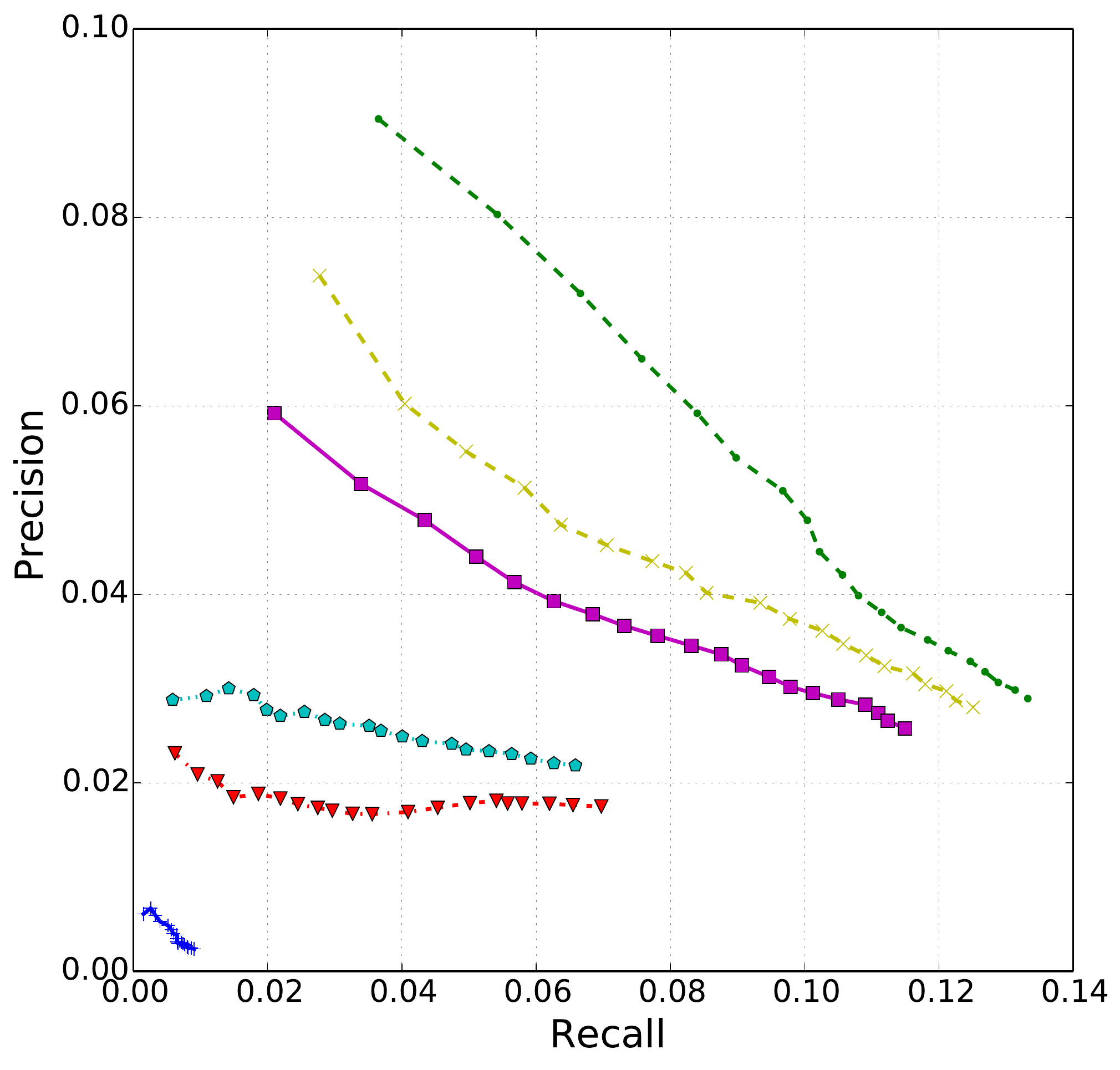} 
   } 
   \subfloat[Delicious]{ 
      \includegraphics[width=0.33\textwidth]{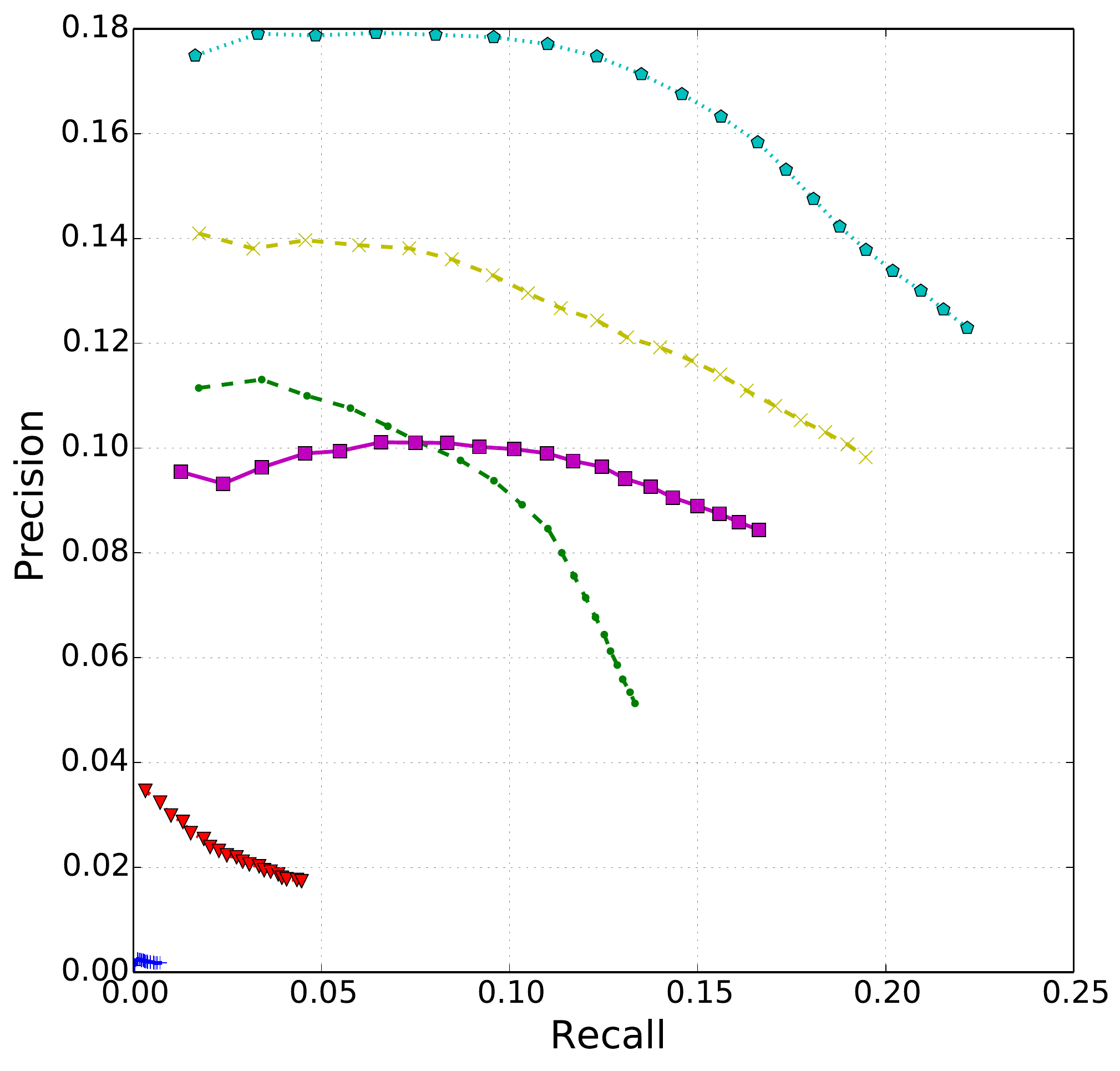} 
   } 
   \caption{Precision / Recall plots for BibSonomy, CiteULike and Delicious showing the recommender accuracy of our approach SUSTAIN+CF$_U$ in comparison to the baseline methods for $k$ = 1 - 20 recommended resources.}
	 \label{fig:prec_rec}
\end{figure*}

A glance at the values in Table \ref{tab:full_norm} and the Precision / Recall plots in Figure \ref{fig:prec_rec} reveals that the simplest baseline algorithm, the unpersonalized MP approach, results in very low estimates of accuracy. This indicates the difficulty of the task of predicting users' resource choices in our three datasets gathered from the social tagging systems BibSonomy, CiteULike and Delicious. However, across all datasets, the remaining personalized algorithms reach larger estimates and therefore, are successful in explaining a substantial amount of variance in user behavior.
 
Referring to our research question \textit{RQ} stated in Section \ref{sec:intro}, as to whether SUSTAIN+CF$_{U}$ can improve CF$_{U}$ in terms of recommender accuracy, our hybrid approach outperforms CF$_{U}$ with respect to all metrics applied and across all three datasets. For instance, in the Precision / Recall plots in Figure \ref{fig:prec_rec}, we can see no overlap between the two corresponding curves -- with SUSTAIN+CF$_{U}$ always reaching higher values than CF$_{U}$. Moreover, especially the results of the ranking-dependent metric nDCG@20 in Table \ref{tab:full_norm} shows a large difference between SUSTAIN+CF$_{U}$ and CF$_{U}$, indicating that our approach can be used to successfully re-rank the candidate resources identified by CF$_{U}$ in a more personalized manner. Hence, we can answer our RQ in the affirmative.
 
Interestingly, the results also clearly show that the performance of the algorithms strongly varies across BibSonomy, CiteULike and Delicious and that in each of the three datasets a different algorithm performs best. For instance, in the case of CiteULike, the best results are reached by CF$_{R}$. We relate this to the fact that the average topic similarity per user in CiteULike (18.9\%) is much higher than in BibSonomy (7.7\%) and Delicious (4.5\%), indicating a more thematically consistent resource search behavior. The higher consistency is in advantage for predictions that are based on resources collected in the past, such as CF$_R$-based predictions. In case of Delicious, the users in the dataset are chosen through a so-called mutual-fan crawling strategy (see \cite{Cantador:RecSys2011}) and thus, are not independent from each other. This is conducive to methods that capture relations between users with common resources by means of high-dimensional arrays, such as WRMF. However, compared to the other algorithms, especially to CF$_{R}$ and WRMF, SUSTAIN+CF$_{U}$ results in relatively robust estimates and even outperforms all other algorithms in case of BibSonomy.

\section{Conclusions and Future Work} \label{sec:conclusion}
With this work we show that a connectionist model of human category learning applied to mimic a user's attentional focus and interpretation helps to improve user-based CF predictions. 
We attribute this improvement to the well established, cognitive plausibility of SUSTAIN \cite{love2004sustain} by which in our case a larger amount of variance in a user's decision making on a given 
set of resources can be explained: Reconstructing the user history in form of an iteratively trained network with history-specific patterns of attentional tunings and clusters, does more justice to a 
user's individuality than a CF-based representation of user-resource relations. Additionally, we observe that our hybrid SUSTAIN+CF$_U$ model is more robust in terms of accuracy estimates but also less complex in 
terms of computational effort than WRMF.

In future, we aim at improving and further validating our model. First, we are working on a variant that is independent of CF$_U$ and searches for user-specific recommendations only by means of the correspondingly 
trained SUSTAIN network. Second, we will make use of the network's sensitivity for a user's mental state to realize a more dynamic recommendation logic. In particular, based on creative cognition
research (e.g., \cite{finke1992creative}), we assume a broader attentional focus to be associated with a stronger orientation toward novel 
resources. Recommendations should become more accurate if this kind of association turns out to be reliable and is integrated into the algorithm.

With respect to evaluation, we regard algorithmic models, such as SUSTAIN, not only as tools to improve recommendations but also to derive specific research questions in the field of Web Science. For instance, we can raise the question whether the extent of a user's curiosity is related to her ability to discover high quality resources before other users. One way to tackle this question would be to draw on the established SPEAR algorithm \cite{yeung2011spear}, differentiating between discoverers and followers and relating its scores with SUSTAIN-specific scores of curiosity. 

We conclude that our attempt to keep the translation from theory into technology as direct as possible holds advantages for both technical and conceptual aspects of recommender research. By applying computational 
models of human cognition, we can improve the performance of existing recommender mechanisms and at the same, gain a deeper understanding of fine-grained level dynamics in a Social Information System.

\section{Acknowledgments}
The authors would like to thank Emanuel Lacic and Sebastian Dennerlein for many valuable comments on this work. This work is supported by the Know-Center, the EU funded projects 
Learning Layers (Grant Agreement 318209) and weSPOT (Grant Agreement 318499) and the Austrian Science Fund (FWF): P 25593-G22. The Know-Center is funded within the Austrian COMET 
Program - Competence Centers for Excellent Technologies - under the auspices of the Austrian Ministry of Transport, Innovation and Technology, the Austrian Ministry of Economics and Labor 
and by the State of Styria. COMET is managed by the Austrian Research Promotion Agency (FFG).

\bibliographystyle{abbrv}
\bibliography{sustain}
\end{document}